\documentclass[prd,aps,psfig,preprint]{revtex4}

\usepackage{graphics}
\usepackage[dvips]{color}
\usepackage{epsfig}
\newcommand{\Ref}[1]{(\ref{#1})}

\begin{document}
\title{Self-wiring in neural nets of point-like cortical neurons fails
 to reproduce cytoarchitectural differences}
\author{Fail M. Gafarov}

\address{Department of Theoretical Physics, \\ Tatar State  University of Humanity and Pedagogic, \\
420021 Kazan, Tatarstan Street, 1 Russia \\
\email*{fgafarov@yandex.ru, fail@kazan-spu.ru}}

\begin{abstract}

We propose a model for description of activity-dependent evolution
and self-wiring between binary neurons. Specifically, this model
can be used for investigation of growth of neuronal connectivity
in the developing neocortex. By using computational simulations
with appropriate training pattern sequences, we show that
long-term memory can be encoded in neuronal connectivity and that
the external stimulations form part of the functioning neocortical
circuit. It is proposed that such binary neuron representations of
point-like cortical neurons fail to reproduce cytoarchitectural
differences of the neocortical organization, which has
implications for inadequacies of compartmental models.

\end{abstract}

\maketitle
\section{Introduction}

The nature of processes, underlying the brain's self-organization
into functioning neocortical circuits remains unclear. For
theoretical description of activity-independent and
activity-dependent emergence of complex spatio-temporal patterns,
and establishment of neuronal connections, various methods have
been proposed \cite{21,22,23,2,5,49}. These theoretical
investigations give a fresh understanding of the observed
phenomena and stimulate discovery of emergent features of neural
circuits. In this paper, we propose a new theoretical approach for
description of activity-dependent evolution and self-wiring
between binary neurons. We propose a model based on the following
experimental data (1)-(3) and hypotheses (4)-(5):
\begin{enumerate}
\item Development of neuronal connectivity dependents on the neurons
activity \cite{46,26,36,1};
\item Direction of motion of the growth cones is controlled by diffusible chemicals -axon guidance molecules
(AGM) \cite{19,8,15,20};
\item Axon's growth rate dependence on the neuron's activity \cite{29,31,30,28};
\item Depolarization causes neurons to release axon guidance  \cite{35,34};
\item  Type of neuronal connectivity is determined postsynaptically during synaptogenesis
\cite{33,40,41}.
\end{enumerate}

A state of all cells at some moment we shall call the activity
pattern. Sequential alternation of activity patterns we shall call
activity dynamics. We suppose that each neuron receives an external
signal from a sensory cell (optical, taste, smell, ets.). Patterns
of external signals we will call training patterns. We demonstrate an example of training process by
stimulation of the developing net by the sequence of training
patterns. The sequence of training patterns can be considered as a
training program. The goal of the training is creation of complex
connections structure between initially disconnected neurons. For mathematical
description of net's electrical activity we use the simplest model \cite{50}.
We take into account only properties which are most important for
description of the activity-dependent self-wiring.

\section{Mathematical framework}

In this section we consider a mathematical basis for description of
the activity - dependent self-wiring in neural nets. For
mathematical description of neurons states and activity patterns, we
use simple representation from modern neurobiology and artificial
neural nets (ANN) theory. At state of rest neuronal
membrane is polarized. When the membrane is locally depolarized up
to the certain value of transmembrane potential by opening the
ligand activated synaptic ion channels, then the voltage gated ion
channels are opened which causes the generation of the action
potential. The action potential travels along the axon and causes
local depolarization of another neurons through synapses. We suppose
that each cell at each moment of time can be in two states: active,
$S_i(t_j)=1$, or inactive, $S_i(t_j)=0$. For real neurons the active
state can be regarded as the real neuron's single spike or a
sequence of spikes with frequency above some value.

We consider the $i$-th neuron's cell body as a spherical object with
center at the point with the radius-vector ${\bf r}_i$ in
extracellular environment. For implementation activity-dependent
release of AGM in our model, we suppose
that neurons release AGM at a moment of depolarization. For
simplification we assume that all neurons fire and release AGM
synchronous, depending on their state. AGM which have released by
neurons diffuse through the extracellular space, then bind axon's
growth cones receptors and control their growth. When we consider
AGM release by neurons, we neglect geometrical properties of them
and consider them as a point sources of AGM.

According to our model if the cell is in active state, $S_i(t_k)=1$,
it releases some amount of AGM. We suppose that all neurons release
the unit amount of the one type AGM which causes only attraction of
growth cones.

For description of AGM diffusion process we use a simple diffusion
equation. The concentration of AGM, $c_{ij}$, released by the i-th
cell at the moment $t_j$ can be found as the solution of the
equation
\begin{equation}
\frac{\partial c_{ij}}{\partial{t}}=D^2 \Delta c_{ij}-k c_{ij}
\end{equation}
with the initial conditions $c_{ij}({\bf r},{\bf r}_i,
t_j)=\delta({\bf r}-{\bf r}_i)S_i(t_j)$. Here $D$ and $k$ are AGM
diffusion and degradation coefficients in the intracellular medium.
We consider here the case without boundary conditions. The solution
of this equation, describing the concentration of AGM at the point
${\bf r}$ at the time $t$ is
\begin{equation}\label{cij}
c_{ij}({\bf r},{\bf r}_i, t, t_j)=\frac{S_i(t_j) }{(2 D \sqrt{\pi
(t-t_j)})^{3}} \exp \left(-k(t-t_j)-\frac{|{\bf r}-{\bf r}_i|^2}{4D^2 (t-t_j)}\right).
\end{equation}
Total concentration of AGM at the point $\mathbf{r}$ can be found by
summation of  concentrations of AGM which were released by each cell
\begin{equation}
C({\bf r},t)=\sum_{i=1}^{N} \sum_{j=1}^k c_{ij}({\bf r},{\bf
r}_i, t, t_j). \label{f3}
\end{equation}

According to the experimental data we suppose that a growth cone
will move only if its soma is at inactive state, and the force
acting on it is proportional to AGM concentration gradient $\nabla
C$ at the growth cone's position. Therefore, the equation of motion
of the $n$-th neuron's $k$-th growth cone, described by the
radius-vector ${\bf g}_k^n$ in the chemical field  can be written in
the following form
\begin{equation}
\frac{d {\bf g}_k^n}{dt}=\lambda \nabla C({\bf g}_k^n,t)
[S_n(t)-1]. \label{f4}
\end{equation}
Here $\lambda$ is a coefficient describing axon's sensitivity and
motility. Taking into account the expression for total concentration
from Eqn. \Ref{f3} and using it into Eqn. \Ref{f4} we obtain
\begin{equation}
\frac{d {\bf g}_k^n}{dt}=\lambda [S_n(t)-1] \sum_{i=1}^{N}
\sum_{j=1}^k \nabla c_{ij}({\bf g}_k^n,{\bf r}_i, t, t_j).
\label{f4}
\end{equation}

From Eq. \Ref{cij} we find the expression for gradient of AGM
concentration $c_{ij}$ in the form below
\begin{equation}
\nabla c_{i,j}({\bf r},{\bf r}_i,t,t_i)= -\frac{S_i(t_j)({\bf
r}-{\bf r}_i)}{16 D^5 \pi^{\frac{3}{2}} (t-t_j) ^\frac{5}{2}} \exp
\left(-k(t-t_j)-\frac{|{\bf r}-{\bf r}_i|^2}{4D^2(t-t_j)}\right). \label{f5}
\end{equation}

If  a some growth cone is close to the another cell's soma, i.e if
$|{\bf g}_k^n-{\bf r}_i|<\varepsilon$ ($\varepsilon$ can be
considered as the soma's geometrical radius) then synaptogenesis
process takes place, and synaptic connection between these neurons
will be established. Type of the neuronal connections between $i$-th
and $n$-th neurons we describe by using synaptic weights $w_{n,i}$
($w_{n,i}=1$ means  excitatory and $w_{n,i}=-1$ inhibitory
connections). In our model the type of synaptic connection
established between neurons depends on the state of postsynaptic cell
at synaptogenesis moment (if $S_i(t_k)=1$ then $w_{n,i}=1$, else
$w_{n,i}=-1$).
\begin{figure}[bt]
\centerline{\psfig{file=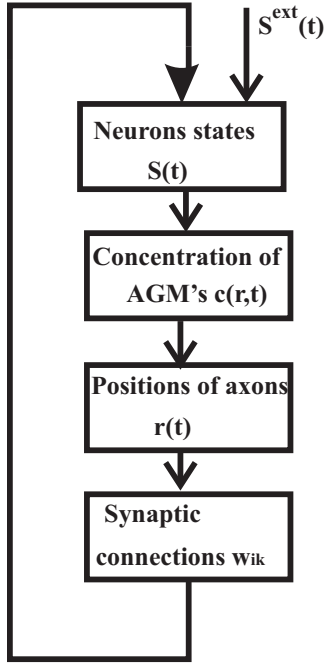,width=2in}}
\caption{A flowchart of self-wiring process}
\end{figure}
For implementation network's electrical activity dynamics several
models can be used (firing rate models, integrate and fire neurons,
etc.). For simplicity we will exploit the simplest one. We assume
that the state of each cell at next moment of time is determined by
states of another neuron, and neuronal connections weights, and
external signal. Each postsynaptic cell integrates inputs coming
from all presynaptic neuron and an external signal $S_i^{ext}$
\begin{equation}
s=\sum_{n=1}^N w_{n,i} S_n(t_i)+S_i^{ext},
\end{equation}
and the state of each neuron at next moment is determined by
following rule: if $s>0$ then $S_i(t_{i+1})=1$, else
$S_i(t_{i+1})=0$.

This  model gives a closed set of equations describing AGM's release
and diffusion, and axons grow and neuronal connections establishment
as well as the net's electrical activity dynamics. In the flowchart
(Fig. 1) the back loop in the neural net's activity-dependent
development and self-wiring are shown. The concentration of AGM in
the extracellular space is controlled by neurons state. Growth and
movement of growth cones is managed by the concentration gradients
of AGM. Growth cones can make neuronal connections with other
neurons and change the network's connections structure which change
the network's activity.

\section{Numerical analysis}

In this section we present the result of the numerical simulation of
the self-wiring net presented above. For testing possibilities and
properties of self-wiring nets we use the net with $27$ neurons,
$N=27$, placed at the points with radius-vectors ${\bf
r_1}=(0,0,1)$, ${\bf r}_2=(0,0,0)$, ${\bf r}_3=(0,-1,0)$, ... ${\bf
r}_{27}=(1,1,-1)$ (see Fig. 3). Each neuron has $27$ growth cones,
which can be considered as branches of its axon. Initially all growth
cones are located near the soma ${\bf g}_k^n={\bf r}_k+{\bf
\varepsilon}$, $|{\bf \varepsilon}|<0.01$, and all synaptic weights
equal to zero ($w_{ik}=0$ $i,j=1...N$).

We can write the Eqn. \Ref{f5} for AGM's gradient using a discrete
time $t=n \Delta t$ (present time), $t_m= m \Delta t$ (release time)
in the following form
\begin{equation}
\nabla c({\bf r},{\bf r}_i,m\Delta t,n\Delta t)= \frac{S_i(n\Delta
t)({\bf r}-{\bf r}_i)}{16 D^5 \pi^{\frac{3}{2}} ((n-m)\Delta
t)^\frac{5}{2}} \exp \left(-kt-\frac{|{\bf r}-{\bf
r}_i|^2}{4D^2(n-m)\Delta t}\right), \label{f6}
\end{equation}
where $\Delta t$ is a time interval between the net's iterations.
For simplification we set $\Delta t=1$ and rewrite the Eqn. \Ref{f6}
in terms of indices $m$, $n$
\begin{equation}
\nabla c_{m,n}({\bf r},{\bf r}_i)= \frac{S_i(n)({\bf r}-{\bf
r}_i)}{16 D^5 \pi^{\frac{3}{2}} (n-m)^\frac{5}{2}} \exp
\left(-kt-\frac{|{\bf r}-{\bf r}_i|^2}{4D^2(n-m)}\right).
\label{f7}
\end{equation}
\begin{figure}[bt]
\centerline{\psfig{file=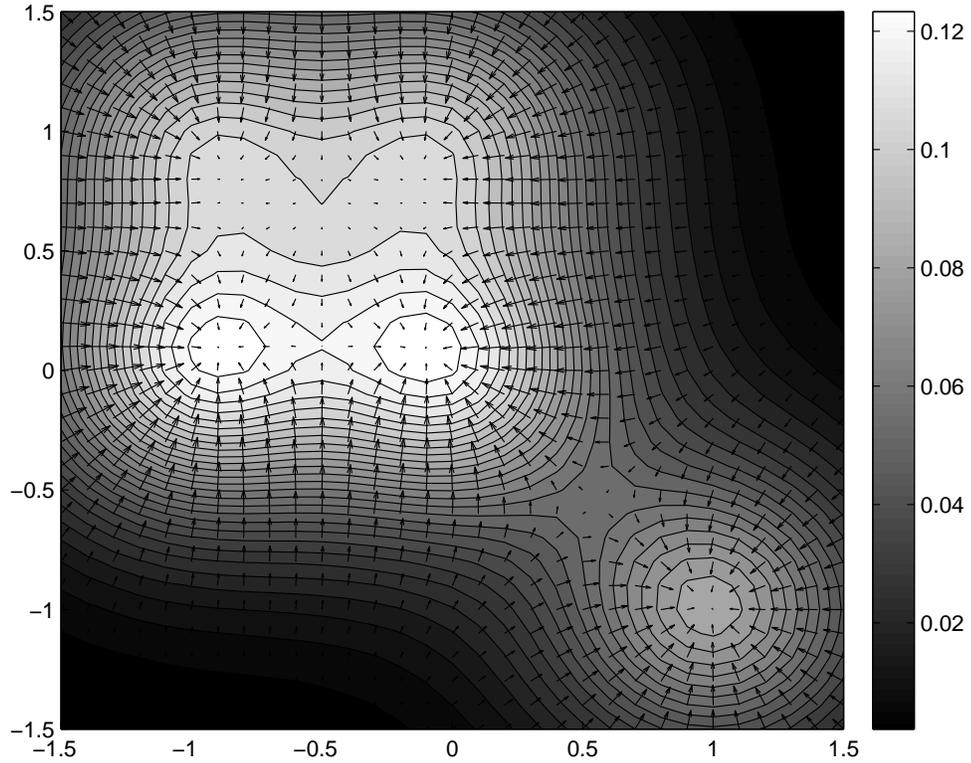,width=5.5in}}
\caption{The contour plot of target-derived AGM concentration. The concentration of AGM is higher in lighter regions of the figure.
The arrows reproduce the concentration gradients which means the axons movement
directions for the particular place.}
\end{figure}
For numerical integration a growth cone equation of motion \Ref{f4}
we have to rewrite it in terms of the discrete time. Using the time
interval $\Delta t=1$ we obtain the recursion relation which allow
us calculate the spatial position of each growth cone at the moment
$k+1$ in terms of the $k$-th moment
\begin{equation}
{\bf g}_{k+1}={\bf g}_k+\lambda[S_n(t_k)-1] \sum_{i=1}^N
\sum_{m=1}^k \nabla c_{m,k}({\bf g_k},{\bf r}_i).
\end{equation}

In above equation we have omitted indices, describing a cell's and
the growth cone's number. The value $\Delta t=1$ is not critical for
numerical calculations of the model and it was used only for
simplification point of view. Different values of parameters $D$,
$k$, $\lambda$ gives different connectivity patterns between
neurons, because these parameters characterize growth cone's
movement speed, and AGM's acting distance, and etc. Basic properties
of the net have remained unchanged. For all subsequent numerical
calculations we used the following values: $D=0.1$,
$k=0.1$, $\lambda=10$.

In Fig. 2 we present the contour plot of the section by plane
$z=0.02$ the static AGM concentration when the $6$-th, $7$-th,
$15$-th, $16$-th, $19$-th neurons are in active state. The arrows
reproduce the concentration gradients which means the axons movement
directions for the particular place.
\begin{figure}[bt]
\centerline{\psfig{file=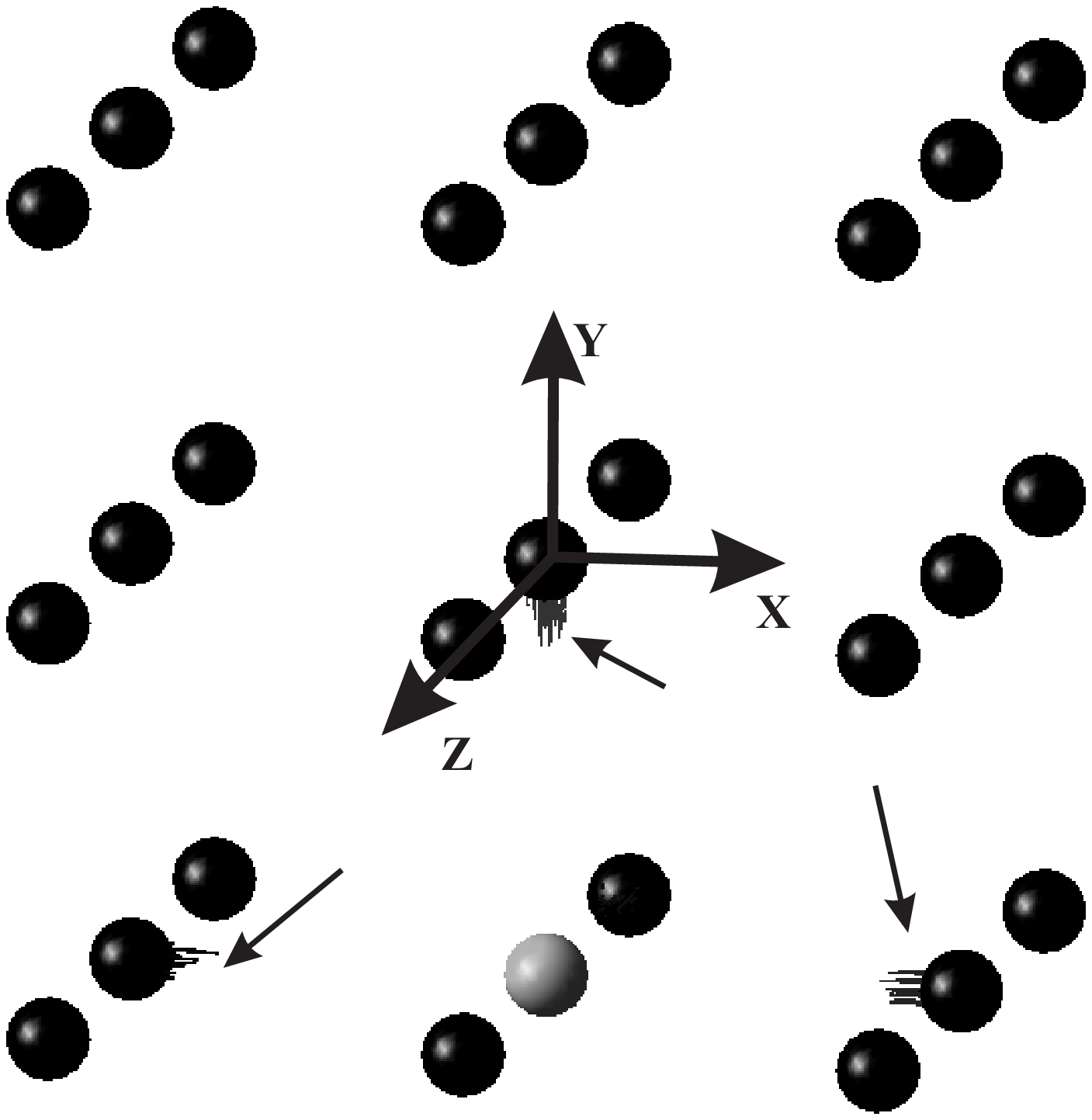,width=5.5in,angle=0}}
\caption{Activity-dependent evolution at the beginning of simulation $(n=110)$, see Section III.
One of the cell has stimulated by the external signal $S_{11}^{ext}=1$ (bright sphere),
and other neurons are at inactive state. The axon's branches of the nearest neurons
have started to grow in the direction of the active cell because it releases AGM.}
\end{figure}
\begin{figure}[bt]
\centerline{\psfig{file=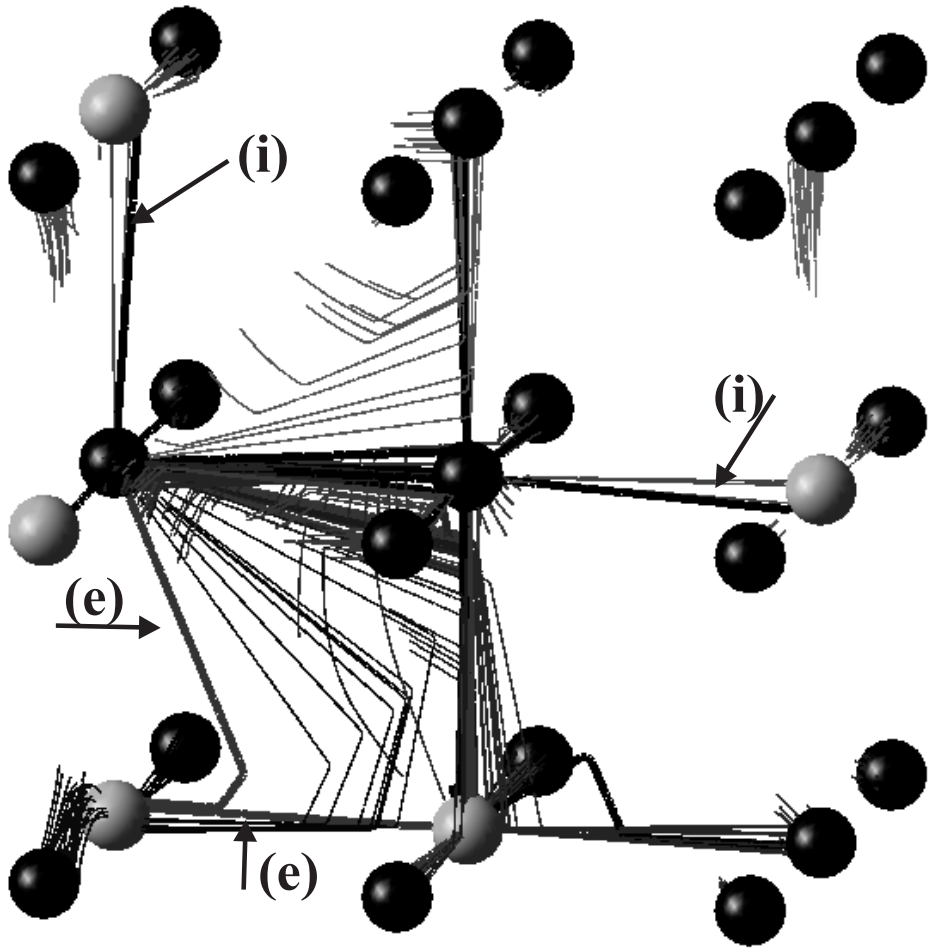,width=5.5in,angle=0}}
\caption{Activity-dependent evolution at $n=680$. Between some of the neurons inhibitory (i) and excitatory (e)
connections have been established.}
\end{figure}
\begin{figure}[bt]
\centerline{\psfig{file=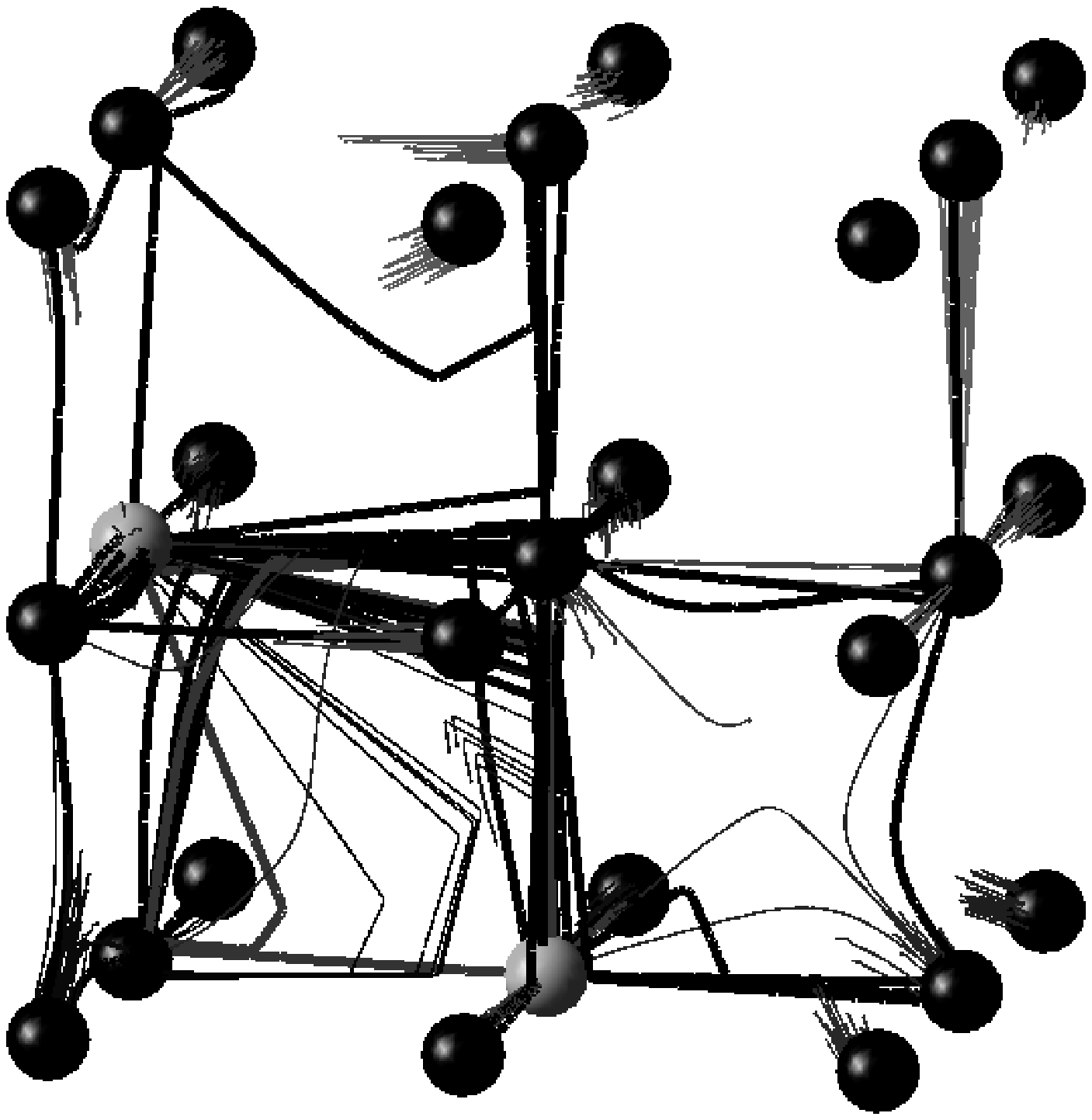,width=5.5in,angle=0}}
\caption{Activity-dependent evolution at $(n=910)$. By using the training program  the complicated
connections pattern between neurons have been created. The network has following neuronal connections:
$w_{10,11}=1$, $w_{12,11}=1$ ,$w_{2,11}=1$, $w_{13,14}=-1$,
$w_{15,14}=-1$, $w_{5,14}=-1$, $w_{23,14}=-1$, $w_{17,14}=-1$,
$w_{11,14}=1$, $w_{6,5}=-1$, $w_{4,5}=-1$, $w_{8,5}=-1$,
$w_{2,5}=1$, $w_{11,5}=-1$, $w_{17,5}=-1$, $w_{15,5}=-1$,
$w_{13,5}=-1$, $w_{12,5}=1$, $w_{14,5}=1$, $w_{10,5}=1$,
$w_{14,11}=-1$, $w_{20,11}=-1$, $w_{1,2}=-1$, $w_{3,2}=-1$,
$w_{7,8}=-1$, $w_{23,24}=-1$, $w_{22,23}=-1$, $w_{26,23}=-1$,
$w_{9,6}=-1$, $w_{15,6}=-1$, $w_{9,8}=-1$, $w_{17,8}=-1$,
$w_{3,6}=-1$, $w_{12,2}=-1$, $w_{14,2}=-1$, $w_{5,11}=-1$,
$w_{5,6}=-1$, $w_{14,23}=-1$, $w_{17,6}=-1$, $w_{12,6}=-1$,
$w_{14,6}=-1$, $w_{10,6}=-1$, $w_{20,6}=-1$, $w_{20,2}=-1$,
$w_{13,16}=-1$, $w_{20,23}=-1$, $w_{10,2}=-1$, $w_{13,11}=-1$,
$w_{5,2}=1$.}
\end{figure}

By using the appropriate training patterns sequence the net is built
with inhibitory and excitatory connections which is evolved in
oscillatory manner without any external stimulation. As an example
we present the results of the network training process which was obtained
by using the following training pattern sequence: $S_{11}^{ext}=1$
for $10<n<192$; $S_{14}^{ext}=1$ for $192<n<360$; $S_{5}^{ext}=1$
for $360<n<515$; $S_{5}^{ext}=1$ for $525<n<576$; $S_{2}^{ext}=1$
for $580<n<875$; $S_{6}^{ext}=1$ for $580<n<880$; $S_{8}^{ext}=1$
for $580<n<880$; $S_{23}^{ext}=1$ for $580<n<880$. In Fig. 3 we
reproduce three dimensional picture of the state of the net at the
simulation beginning. Individual neurons (whilst without neuronal connections)
depicted as spheres and its states depicted by color
(bright - active state, dark- inactive). The travelling axon's
branches are depicted as thin curves. One can see from this figure,
how axons grow toward the active cell. When the growth cone reaches
the soma of another neuron, two neurons became connected and axon's
connected branch is depicted by thick curve (Fig. 4). The synapse
type (inhibitory or excitatory) is depicted by color of a curve
(bright - excitatory, dark- inhibitory). When the branch of the same
axon reaches the another cell, which has already connected with it,
this branch will be deleted. The neural net obtained by training is
shown in Fig. 5.

Further training of this net is a hard problem because the internal
oscillation will influence to the training process. The oscillating
patterns of neuronal activity will cause establishment of new
connections, even after switching off the external stimulation, and
can cause uncontrolled self-wiring. In forthcoming papers we plan to
develop programming (training) methods in order to build
the net with initially defined connections.

We can gain access to the patterns stored in the net by stimulating
them by external patterns. Because of an external stimulation will
cause activation of neurons and release of AGM we conclude that access
to the memory can change the connectivity pattern, i.e. patterns
stored in the memory.

\section{Conclusions}

In this paper we presented a mathematical model describing processes
of self-wiring in neural nets. The model can give us a new
understanding of real neural nets development and functioning. By
using the computer simulations we showed that the long-time memory
can be encoded in neuronal connectivity and in what way the external
stimulations builds a functioning neural network.

Hypotheses proposed here should be tested experimentally (in vivo,
in vitro), for verification the possibilities of presented model
application for qualitative and quantitative description of a
neuronal integration processes in a real nets. Our approach gives
only general modeling scheme of activity-dependent network formation
processes and to accordance with experimental data the model
can be complicated and a particular rules describing separate
processes can be changed.

The theoretical framework developed here, can be used for
description the development of a particular set of neurons
constituting  the neural system, which can be located in the media
containing also another type neurons and glia cells. In our model
axon's branches began to growth directly from the soma. In a real
nets growth cones can be guided at the growth beginning  by another
mechanisms (genetically determine contact adhesion, etc) and then by
the mechanism presented in this paper. Geometrical properties and
activity of real neurons are more complicate and they have dendrites.
Here we used the most simple representations of
hypothetical neurons (large spherical soma without dendrites ) and
its activity (binary neurons). Using the approach described here it
is possible to construct a model for description real particular
networks, using more realistic models of neurons (ingrate-and-fire,
Hodgkin-Huxley, etc) and taking into account neurons geometrical
properties and dendrites. We suppose that the model presented
can be used for description self-wiring processes in a developing
(embryonal) as well as adult neuronal systems.

Cytoarchitectural differences between different cortical areas
are the result of differences in distributions of
neuronal phenotypes and morphologies. Self-wiring between binary neurons fails to reproduce cytoarchitectural differences
of the neocortical organization, which has implications for inadequacies of compartmental models.
In order to reproduce cytoarchitectural differences, axonal and dendritic morphologies of single neurons \cite{Kal} need to be integrated into the
model.


\begin{thebibliography}{00}
\bibitem{21} Andras P, A model for emergent complex order in small neural networks, {\it J. Integr. Neurosci.}
{\bf 3}:55-69, 2003
\bibitem{50}Ascoli GA,  Passive dendritic integration heavily affects spiking dynamics of recurrent networks,
 {\it Neural Netw.} {\bf 16}:657-663,2003.
\bibitem{35} Balkoweic A, David MK, Activity-dependent release of endogenous brain-derived neurotrophic
factor from primary sensory neurons detected by ELISA in Situ, {\it J. Neurosc.} {\bf 20}:7417-7423, 2000.
\bibitem{33} Borodinsky LN, Root CM, Cronin JA, Sann SB, Gu X, Spitzer C, Activity-dependent homeostatic specification of transmitter
expression in embrionic neurons, {\it Nature} {\bf 429}:523-530,
2004.
\bibitem{46} Catalano SM Shatz CJ, Activity-Dependent Cortical Target Selection by Thalamic Axons, {\it Science} {\bf 281}:559-562, 1998.
\bibitem{22} Chauvet GA, On the mathematical integration of the nervous tissue based on the S-propagator formalism I: Theory,
{\it J. Integr. Neurosci.} {\bf 1}:31-68, 2002
\bibitem{23} Chauvet P, Chauvet GA, On the mathematical integration of the
nervous tissue based on the S-propagator formalism II: Numerical
simulations for molecular-dependent activity, {\it J. Integr. Neurosci.} {\bf 1}:157-194, 2002.
\bibitem{40} Ciccolini F, Collins TJ, Sudhoelter J, Lipp P, Berridge MJ, Local and Global Spontaneous Calcium Events Regulate
Neurite Outgrowth and Onset of GABAergic Phenotype during Neural Precursor Differentiation, {\it J. Neurosc.} {23}:103-111, 2003.
\bibitem{19} Dickson BJ, Molecular Mechanisms of Axon Guidance, {\it Science} {\bf 298}:1959-1964, 2002.
\bibitem{41} Dugu\'e GP, Dumoulin A, Triller A, Dieudonne S, Target-Dependent Use of Coreleased Inhibitory Transmitters
at Central Synapses, {\it J. Neurosc.} {\bf 25}:6490-6498, 2005.
\bibitem{8}  Goodhill GJ, Gu M, Urbach JS, Predicting axonal response to molecular
gradient with a computational model of filopodial dynamics, {\it Neural Comp.} {\bf 16}:2221-2243, 2004.
\bibitem{29} Gomez TM, Spitzer NC, Regulation of growth cone behavior by calcium: new dynamics to earlier perspectives,
{\it J. Neurobiol}, {\bf 44}:174-183, 2000.
\bibitem{31} Gomez TM, Spitzer NC, In vivo regulation of axon extension and pathfinding by growth-cone calcium transients,
{\it Nature}, {\bf 397}:35-355, 1999.
\bibitem{34} Hartmann M, Heumann R, Lessman V, Synaptic secretion of BDNG after high-frequency stimulation
in glutamatergic synapses, {\it The EMBO J.} {\bf 20}:5887-5897, 2001.
\bibitem{30} Henley J, Poo MM, Guiding neuronal growth cones using $Ca^{2+}$ signals, {\it Trends Cell Biol.},
{\bf 14}:320-330, 2004.
\bibitem{2}  Hentschel HGE, van Ooyen A, Models of axon guidance and bunding during
development, {\it Proc. R. Soc. Lond} {\bf B 266}:2231-2238, 1999.
\bibitem{Kal} Kalisman N, Silberberg G,  Markram H, Deriving physical connectivity from neuronal morphology,
{\it Biol. Cybern.} {\bf 88}:210-218, 2003.
\bibitem{26} Kandler K, Activity-dependent organization of inhibitory circuits:
lessons from the auditory system, {\it Curr. Opin. Neurobiol.} {\bf 14}:96-104, 2004.
\bibitem{28} Ming GL, Henley J, Tessier-Lavigne M, Song H.-J, Poo M.-M, Electrical activity modulates growth cone
guidance by diffusible factors, {\it Neuron}, {\bf 29}:441-452,
2001.
\bibitem{15} Nieto MA, Molecular Biology of Axon Guidance, {\it Neuron} {\bf 17} :1039-1048, 1996.
\bibitem{36} van Ooyen A, van Pelt J, Activity-dependent outgrowth of neurons and overshot phenomena in developing neural networks,
{\it J.theor. Biol} {\bf 167}:27-43,1994.
\bibitem{1}  van Ooyen A, van Pelt J, Complex Periodic Behavior in a
neural network model with activity-dependent neurite outgrowth,
{\it J. Theor. Biol.} {\bf 179}:229-242, 1996.
\bibitem{5} van Pelt J, Kamermans M, Levelt CN, Van Ooyen A, Ramakers GJA, Roelfsema PR (eds.),
Development, Dynamics and Pathology of Neuronal Networks: From
Molecules to Functional Circuits, {\it Prog. Brain Res.} {\bf 147}, Elsevier, Amsterdam, 2005.
\bibitem{49} Segev R, Ben-Jacob E, Generic modeling of chemotactic based self-wiring of neural networks, {\it Neural Networks} {\bf 13}:185-199, 2000.
\bibitem{20} Tessier-Lavigne M., Goodman CS, The Molecular Biology of Axon Guidance, {\it Science} {\bf 274}:1123-1133, 1996.

\end{thebibliography}
\end{document}